\documentclass[10pt,letterpaper,twocolumn]{article} 

\usepackage{ol2}
\usepackage[draft,implicit=false]{hyperref}
\usepackage[latin9]{inputenc}
\usepackage{amsmath}
\begin{document}

\twocolumn[ 

\title{A silicon single-crystal cryogenic optical resonator}


\author{Eugen Wiens, Qun-Feng Chen, Ingo Ernsting, Heiko Luckmann, Ulrich
Rosowski, Alexander Nevsky, and Stephan Schiller$^{*}$}

\address{
Institut für Experimentalphysik, Heinrich-Heine-Universität Düsseldorf,
Düsseldorf, Germany \\
$^*$Corresponding author: Step.Schiller@uni-duesseldorf.de
}

\begin{abstract}
We report on the demonstration and characterization of a silicon optical
resonator for laser frequency stabilization, operating in the deep cryogenic
regime at temperatures as low as 1.5~K. Robust operation was achieved,
with absolute frequency drift less than 20 Hz over 1~hour. This stability allowed
sensitive measurements of the resonator thermal expansion coefficient
($\alpha$). We found $\alpha=4.6\times10^{-13}$ ${\rm K^{-1}}$ at
1.6~K. At 16.8~K $\alpha$ vanishes, with a derivative equal
to $-6\times10^{-10}$ ${\rm K}^{-2}$. The temperature of the resonator
was stabilized to a level below 10~$\mu$K for averaging 
times longer than 20~s. The sensitivity of the resonator frequency
to a variation of the laser power was also studied. The corresponding
sensitivities, and the expected Brownian noise indicate that this
system should enable frequency stabilization of lasers at the low-$10^{-17}$
level. 
\end{abstract}
] 
%
Optical resonators with low sensitivity to temperature and mechanical
forces are of significant importance for precision measurements in
the optical and microwave frequency domain. In the optical domain,
they serve to stabilize the frequencies of lasers for spectroscopic
applications, notably for optical atomic clocks, and for probing fundamental
physics issues such as the properties of space-time. Also, by conversion
of ultrastable optical frequencies to the radio-frequency domain via
an optical frequency comb, radio-frequency sources with ultra-low
phase noise can be realized \cite{NISTULEcavity}, leading to e.g.
radar measurements with improved sensitivity.

The conventional approach for ultra-stable optical resonators is the
use of ULE (ultra-low expansion glass) material, operated at temperatures
near room temperature, where the coefficient of thermal expansion
(CTE) exhibits a zero crossing. While ULE resonators with optimized designs
(long length, acceleration-insensitive shape) have reached impressive
performance \cite{Nicholson2012}, their operating temperature near
300~K necessarily leads to a level of Brownian length fluctuations
which imposes a fundamental limit to the achievable frequency stability
\cite{Numata},\cite{Cole}. Cryogenic operation of a resonator provides
one avenue towards reduction of these fluctuations. The Allan deviation
of length fluctuations decreases proportional to $\sqrt{T}$
\cite{Numata}, if the mechanical dissipation of the resonator elements,
in particular of the mirror coatings, is independent of temperature.
Measurements performed thus far indicate that the dissipation
of mirrors with crystalline substrates at cryogenic temperature are
indeed similar to those of fused silica mirrors at room temperature
\cite{Yamamoto},\cite{Granata}. Nowadays, robust cryogenic solutions
exist for continuous operation of  even fairly large objects, such
as optical resonators, at temperatures as low as 0.1~K. This offers
the possibility of reduction of resonator length fluctuations by more
than one order of magnitude compared to today's lowest levels realized
at room temperature, with a corresponding reduction in frequency instability
of the laser stabilized to the resonator. A second outstanding feature
of crystalline cryogenic optical resonators is the absence of length
drift thanks to the near-perfect lattice structure.

Cryogenic optical resonators made of single-crystal sapphire have been
developed early on \cite{Seel} and operated at temperatures as low
as 1.4~K \cite{StorzDisseration}. Extremely small thermal expansion
\cite{Seel} and long-term drift were achieved \cite{Storz}. Such
resonators have been applied for tests of Lorentz invariance \cite{Antonini},
local position invariance \cite{Mueller} and quantum space-time fluctuations
\cite{Schiller}. Silicon, a machinable optical material available
with high purity and large size, having interesting CTE properties
\cite{Lyon1977}, high stiffness and low mechanical dissipation \cite{McGuigan},
has first been used for an optical reference resonator by Richard
and Hamilton \cite{Richard}. Recently, Kessler et al.~\cite{Kessler2012}
developed a laser frequency stabilization system based on a vertically
supported silicon resonator operated at a temperature of zero CTE,
124~K, and achieved a high frequency stability ($1\times10^{-16}$),
less than 40~mHz laser linewidth, and an extremely low long-term drift.

In this work, we present the first silicon optical resonator for absolute
laser frequency stabilization operated at cryogenic temperature and
discuss its thermal properties. We find that they are compatible with
the goal of achieving frequency instability at the $2\times10^{-17}$
level.

Our silicon resonator consists of a cylindrical spacer of 250~mm
length, 70~mm diameter and 15~mm diameter central hole. One mirror
is flat, while the other has 1~m curvature radius. Spacer and mirror
substrates were manufactured from a dislocation-free float-zone silicon
crystal with {[}110{]} orientation along the cylinder axis. The substrates
and spacer faces were super-polished to a residual surface roughness
less than 0.1~nm. The mirror substrates were coated with a high-reflectivity
coating for 1.5~$\mu$m, and optically contacted to the spacer with
the same orientation as in the spacer crystal. By evaluation of a
ring-down of the laser power transmitted through the resonator at
1.5~K, the finesse and the linewidth of the resonator were determined
to be $2\times 10^{5}$ and 3~kHz, respectively. The coupling efficiency into
the resonator is 15\%.

\begin{figure}[htb]
\noindent \centering{}
\includegraphics[width=8.3cm]{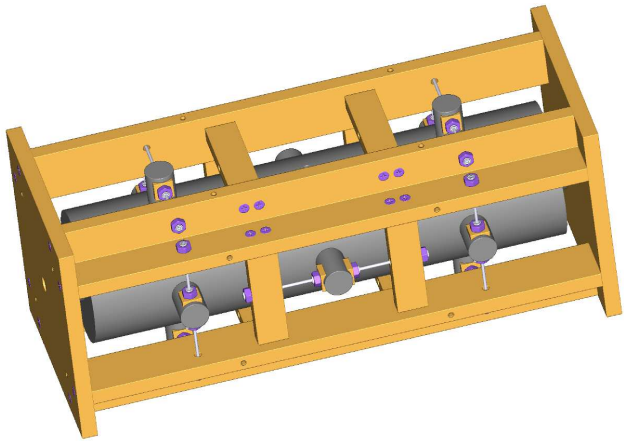}
\caption{Schematic of the copper support structure (yellow) with mounted silicon resonator (grey).}
\label{fig:CavityOverview}
\end{figure}

A special structure for horizontal support of the resonator was implemented (see Fig.~\ref{fig:CavityOverview}).
It was designed for small acceleration sensitivity of the resonator's
length and small stress imposed by the different CTE of resonator
and supporting structure, and for blocking thermal radiation entering
the cryostat through the optical window. The design is based on the
concept developed in Ref. \cite{Sterr}.
The resonator is attached to a copper frame by ten 1~mm
diameter stainless steel wires. Their position was optimized
by the finite-element method (FEM, Comsol) in order to minimize acceleration
sensitivity along the three axes. The estimated sensitivities are 
$3\times10^{-10}/g$ and $3\times10^{-11}/g$ along the resonator optical 
axis and perpendicular to it, respectively. We also simulated the influence of the supporting structure on
the CTE of the resonator. It is a function of temperature, due to the temperature dependence of 
the CTE of both the resonator and the frame. The results indicate negligible dependence of the CTE of the resonator ($\alpha_{reson})$ on a change in frame
temperature at temperatures below 16.8~K. The CTE zero crossing points at 17~K and 124~K 
are shifted by 2~mK and 10~mK, respectively, due to the contribution from the frame.

The frame was rigidly attached to an optical breadboard inside a cryostat.
A 1.5~$\mu$m fiber laser was coupled to the resonator through an
optical window. The frequency was stabilized to a TEM$_{00}$ mode
of the resonator using the Pound-Drewer-Hall (PDH) technique \cite{Drever},
with a standard optical scheme and using a waveguide phase modulator.
The laser power circulating inside the resonator was stabilized by
detection of the light transmitted through the resonator and applying
a feedback signal to an acousto-optical modulator in the optical setup.

The resonator unit was installed inside a low-vibration, two-stage
pulse tube cooler cryostat with Joule-Thomson stage. It allowed a lowest
operating temperature of 1.4~K and a cooling power of 20~mW. Two
calibrated Cernox sensors (inaccuracy less than 10~mK at temperatures
below 30~K) were available for controlling and/or monitoring the
temperature of the supporting breadboard and of the resonator. One
sensor was attached to the center of the breadboard, while the other
was attached to the upper central part of the resonator. An AC resistance
bridge was used to read out the temperature sensors. A resistance
heater installed on the center of the breadboard was used for setting
and maintaining a desired operating temperature. When the temperature
of resonator was stabilized by controlling the resistance of the resonator
sensor, the measured instability at the sensor location was $3.6\times10^{-5}$~K
at $\tau=1$~s averaging time, dropping as $3.6\times10^{-5} K/(\tau/1s)^{1/2}$ for $\tau$ up to $10^4$\,s.
We estimate that the temperature instability of the average temperature
of the resonator is within a moderate factor of the above number, since
the very high thermal diffusivity of silicon at cryogenic temperature
allows for a rapid thermal equilibration within the resonator. The
AC bridge electronics contributes a specified systematic error of
$1.7\times10^{-5}$~K per degree variation of the ambient temperature.
Thus, in a laboratory stabilized to 0.5~K, the resonator temperature
systematic shift is less than $1\times10^{-5}{\rm K}$. 

\begin{figure}[htb]
\noindent \begin{centering}
a)\includegraphics[width=8cm]{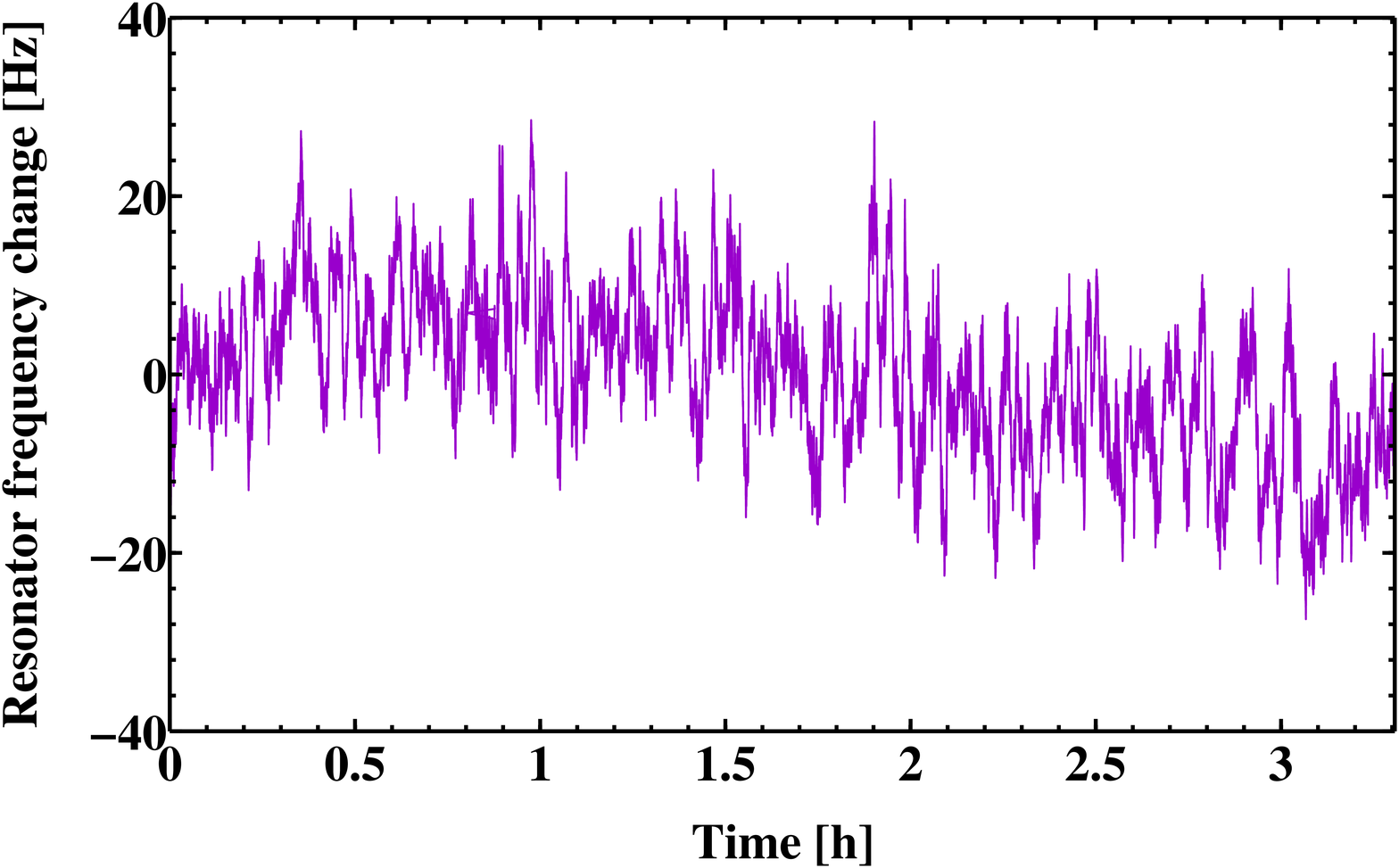}
b)\includegraphics[width=8cm]{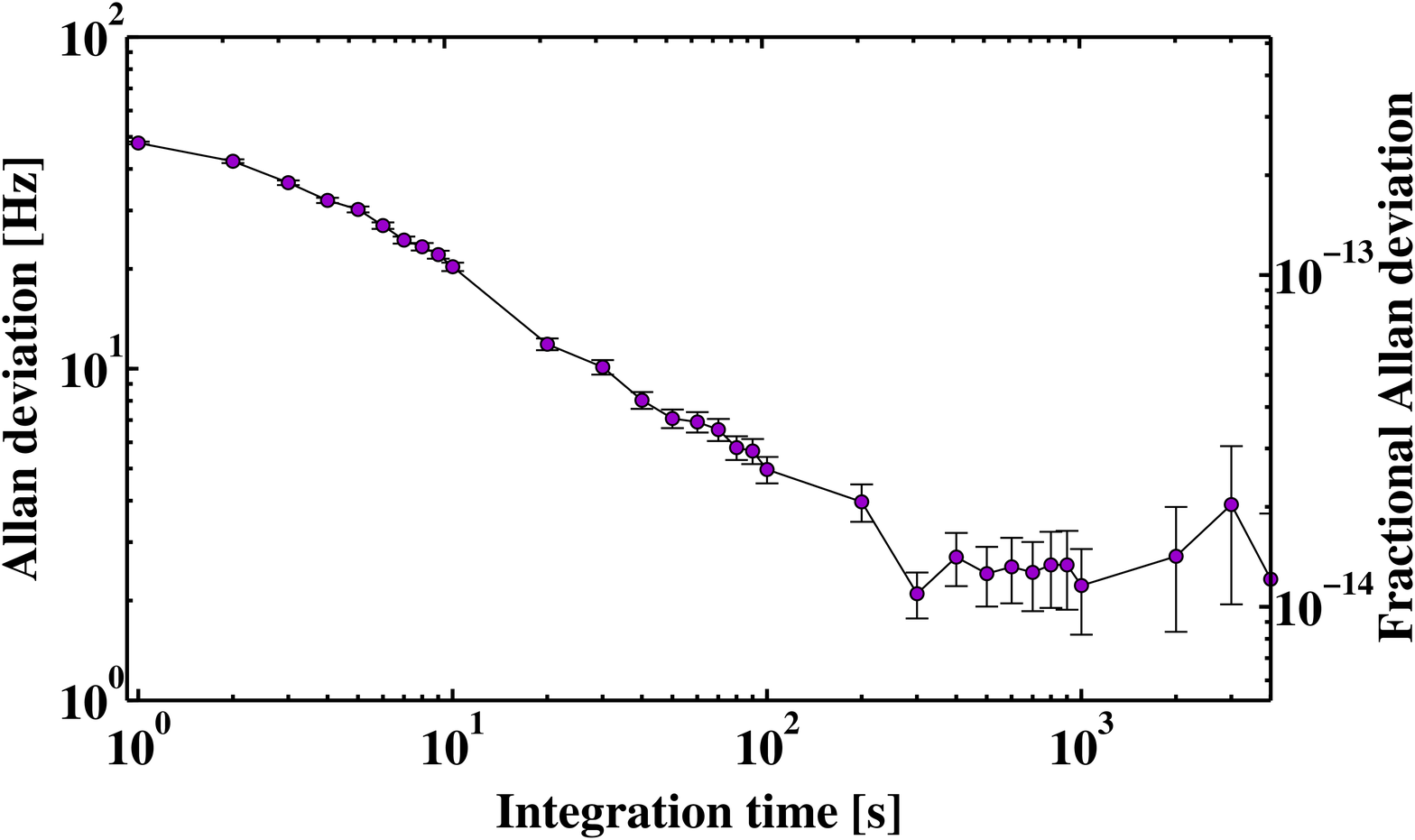}
\par\end{centering}
\caption{a) Time trace averaged over 60~s of a typical resonator frequency measurement relative
to a hydrogen maser, b) corresponding Allan deviation.}
\label{fig:FiguresTypicalTimetrace}
\end{figure}

Measurements of the absolute optical resonator frequency were done
by measuring the frequency of the laser stabilized to it with respect 
to a hydrogen maser, using a commercial erbium-doped fiber laser
frequency comb. A comb mode near 1064~nm was phase-locked to a 1064~nm optical reference
having $2\times10^{-15}$ short-term frequency instability. This reduced significantly 
the linewidth of comb modes in the 1.5~$\mu$m spectral range and permitted 
achieving a beat note between the Si-resonator-stabilized laser and the comb with
a linewidth as low as 1.7~kHz. The laser light was sent from
the laser to the comb via a 150~m long fiber. A typical
resonator frequency time trace and the corresponding instability,
with all parameters including resonator temperature held constant, is depicted
in Fig.~\ref{fig:FiguresTypicalTimetrace}. For integration times
longer than 1000~s the frequency instability was approximately 3~Hz
($2\times10^{-14})$. The contributions to this instability will be subject 
of future investigations.

Given this level of absolute frequency instability and the good resolution
of the temperature sensors, the resonator's thermal expansion coefficient
could be accurately measured. The measurements were done either by heating
the resonator or by letting it cool down to 1.5~K after heating it
up to a desired temperature. The rate of temperature change was approximately
1.5~K/h at temperatures above 1.8~K and much lower for temperatures
below 1.8~K, where it was limited by the cooling capacity of the
cryostat. No significant discrepancies were found between the two measurement
procedures.

\begin{figure}[tbph]
\noindent \begin{centering}
a)\includegraphics[width=8cm]{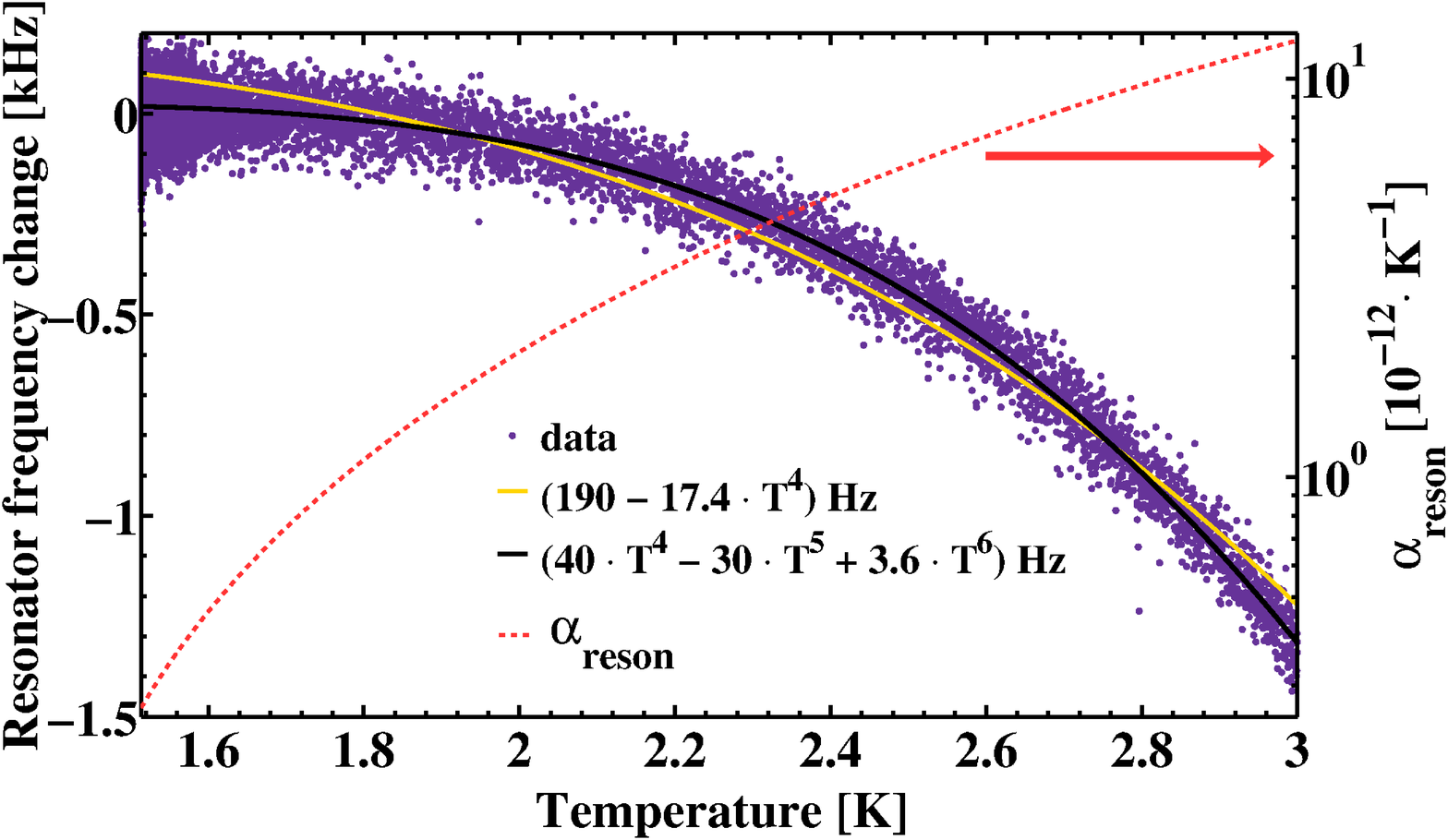}
b)\includegraphics[width=8cm]{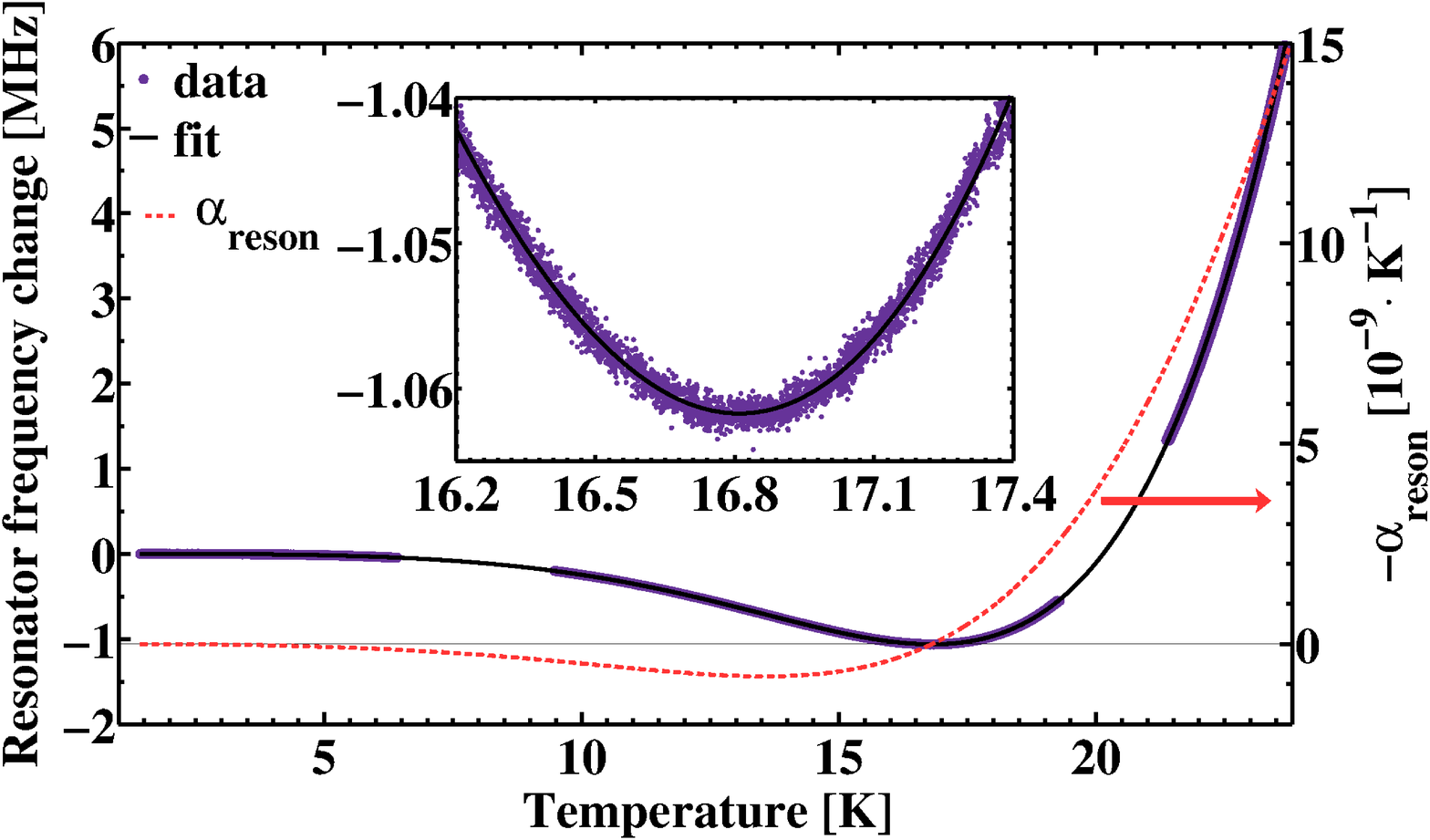}
\par\end{centering}
\noindent \centering{}
\caption{Temperature dependence of the silicon resonator frequency in two temperature
intervals. a) Interval 1.5~K to 3~K, with fit results in the legend
and corresponding expansion coefficient $\alpha_{reson}$ calculated
from the second fit in the legend. b) Interval from 1.5~K to 23.8~K
with a polynomial fit and calculated expansion coefficient
$\alpha_{reson}$. The inset in b) shows a zoom of the interval
where the CTE crosses zero.}
\label{fig:ResultsDifferentRegions}
\end{figure}

The temperature dependence of the resonator frequency is presented
in Fig.~\ref{fig:ResultsDifferentRegions}. The total change in frequency
from 1.5~K to 23.8~K is 6~MHz, with most of the shift occurring
in the region above 20~K. A minimum of the frequency occurs at $T_{\alpha=0}=16.81$~K
with negative frequency derivative $df/dT$ (length expansion) below
this temperature and positive (length contraction) above it \cite{Smith1975}.
Between 1.6~K and 2~K the shift in frequency is only 100~Hz $(5\times10^{-13})$.

Whereas the CTE of an insulating crystal at low temperatures is expected
to be positive and proportional to $T^{3}$ \cite{Lyon1977}, in the
case of silicon its thermal expansion behaviour (expansion turns into
contraction between 16.8~K and 124~K) necessarily implies that such
a dependence can accurately hold only for $T\rightarrow0$. The value
predicted from compressibility and heat capacity measurements is $\alpha_{Si}=$
$4.8\times10^{-13}{\rm K}^{-1}(T/{\rm K})^{3}$ \cite{Lyon1977}.
As displayed in Fig.~\ref{fig:ResultsDifferentRegions}, a pure $T^{3}$
dependence does not fit the data well. Therefore, the data were fitted
with a function containing additional terms. At the lowest
temperatures, we find $\alpha_{reson}(T\rightarrow0)=
(1.1 T^{3}-7.8 T^{4}+8.4 T^{5})\times10^{-13}\,{\rm K^{-1}}$,
where $T$ is the temperature in Kelvin. Our
measured $\alpha_{reson}$ is thus lower then the predicted value
for bulk silicon at temperatures below 3~K. At 1.6~K, the value is $\alpha_{reson}=4.6\times10^{-13}/{\rm K}$.

The frequency data around the frequency minimum at $T_{\alpha=0}=16.8\,{\rm K}$
were accurately fitted by a fourth-order polynomial from which we
obtain the resonator expansion coefficient $\alpha_{reson}(T\simeq T_{\alpha=0})=\sum_{n=0}^{2}B_{n}(T/{\rm K})^{n}$
with $B_{n}=(37\times10^{-9},$ $-4.94\times10^{-9},$ $1.64\times10^{-10})\,{\rm K^{-1}}$.
The complete dataset from 1.6~K to 23.8~K was fitted with a ninth-order
polynomial (Fig.~\ref{fig:ResultsDifferentRegions}b),
from which we obtain the resonator thermal expansion coefficient $\alpha_{reson}(T)$
shown in Fig.~\ref{fig:ResultsDifferentRegions}~b).
It is given by $\alpha_{reson}(T)=\sum_{n=0}^{8}A_{n}(T/K)^{n},$
with $A_{n}=$ $(-9\times10^{-22},\,4\times10^{-20},\,-7\times10^{-19},\,2\times10^{-17},\,-3\times10^{-16},\,2\times10^{-15},\,-1\times10^{-14},\,3\times10^{-14},\,-2\times10^{-14})\,{\rm K^{-1}}$.

To our knowledge, only two measurements of thermal expansion of silicon
at cryogenic temperature, down to 12~K and 6~K \cite{Lyon1977}
were performed previously. However, due to the difficulty of measuring
the corresponding small CTEs, accurate CTE values were only given
above 13~K \cite{Lyon1977}, where they are larger than $1\times10^{-9}/{\rm K}$.
Our value of $T_{\alpha_{reson}=0}=16.8\,{\rm K}$ is similar, but
not identical to that of Ref.~\cite{Lyon 1977}, where 17.8~K was
measured. Our expansivity temperature derivative at 16.8~K is $d\alpha_{reson}(T_{\alpha_{reson}=0})/dT=-6.0\times10^{-10}/{\rm K^{2}}$,
which is comparable to $-8.9\times10^{-10}/{\rm K^{2}}$ at 17.8~K
in Ref.~\cite{Lyon 1977}. This expansivity derivative at 16.8~K
is nearly a factor 20 smaller than the one at the second zero-CTE temperature
124.2~K, $1.7\times10^{-8}/{\rm K^{2}}$ \cite{Kessler2012}.
This means that 16.8~K could also be a candidate operating temperature,
with the advantage of less temperature sensitivity, possibly smaller
thermal noise, as compared to 124~K, and less stringent requirements on cryostat construction. 
As mentioned above, the mirror coating dissipation at this temperature is similar
to the room-temperature values for conventional mirror substrates.

With an estimated upper limit of $1\times10^{-4}$~K resonator temperature
instability at 1~s (discussed above) and the measured thermal expansion
$\alpha_{reson}(T=1.6\,{\rm K})=4.6\times10^{-13}\,{\rm K^{-1}}$,
the corresponding fractional frequency instability is about $5\times10^{-17}$
at 1~s and drops to $1\times10^{-17}$ at $\tau=30\,$s. The fractional
frequency shift error due to AC bridge systematics is expected
to be less than $1\times10^{-17}.$


Fluctuations of the laser power incident on the resonator introduce
changes in the power dissipated on the resonator mirrors and result
in unwanted fluctuations in resonator length due to the thermal expansion
of the mirrors and of the spacer. We measured the resonator frequency
change caused by a varying level of intra-resonator power at two different
operating temperatures. We could not observe any effect, setting an
upper limit of 20~Hz change for a 30$\%$ power change, both at 1.6~K
and at 16.8~K. This upper limit corresponds to approximately 
$3\times10^{-14}/\mu{\rm W}$. 

Because of the absence of measurable effect, we also simulated the
effect of laser heating using FEM and found a 1.4~fm mirror distance
change for 10~$\mu$W power dissipated on each mirror at 1.5~K,
i.e. a fractional resonator frequency change of $3\times10^{-16}/\mu{\rm W}$
total dissipated power. The simulation also shows that thermal equilibrium
of the heated mirrors and the spacer is reached with a time constant
of approximately 1~s in case of good thermal contact with the breadboard.
For our experiment, the power level dissipated in the resonator is
estimated as 1.5~$\mu$W. If the transmitted laser power is actively
stabilized to 1\% or better, a level that is expected to be well feasible
using an advanced photodetector, the corresponding fractional frequency
fluctuations would be $5\times10^{-18}$ or less.

In summary, we demonstrated a silicon optical resonator for laser
frequency stabilization that can be operated at temperatures between
1.5~K and 24~K and investigated its thermal properties. Stable,
long-term operation was achieved at temperature as low as 1.4~K.
The support structure for the resonator allows to take advantage of
the low CTE of bulk silicon. The thermal expansion measurement was
performed both at the lowest absolute temperature and with the highest
sensitivity for a silicon object so far. The influence of finite thermal
expansion coefficient and residual temperature instability on the
cavity frequency was determined to be at the $5\times10^{-17}$ fractional
level and below, depending on averaging time. The sensitivity to changes
in circulating laser power is expected to be controllable to the level
better than $1\times10^{-17}$ using an appropriate power stabilization
unit. For this resonator, at the temperature 1.6~K, the expected
total Brownian noise-induced frequency instability is calculated to
be $6\times10^{-18}$, assuming a coating loss angle $\varphi=1$~mrad,
as determined for silicon mirrors at 20~K in Ref.~\cite{Granata}. 

We conclude that the thermal properties of the described system should
allow to stabilize the frequency of a laser to an instability of less
than $2\times10^{-17}$ for integration times larger than 30~s, taking
into account that the instability arising from the laser frequency
locking system itself can be reduced to the level of $1\times10^{-17}$ \cite{Chen2012}.

We are very grateful to Timo Müller (Wacker Chemitronic) for providing
the crystal and to T. Legero and U. Sterr (PTB) for important help
on the resonator. This work has been funded by ESA project no.~4000103508/11/D/JR.
We thank I. Zayer and J. de Vicente for support.


\begin{thebibliography}{99}

\bibitem{NISTULEcavity} Y. Y. Jiang, A. D. Ludlow, N. D. Lemke, R. W. Fox, J. A. Sherman, L.-S. Ma, and C. W. Oates, Nat. Phot. \textbf{5,} 158--161 (2011).

\bibitem{Nicholson2012} T. L. Nicholson, M. J. Martin, J. R. Williams, B. J. Bloom, 
M. Bishof, M. D. Swallows, S. L. Campbell, and J. Ye, Phys. Rev. Lett. \textbf{109,} 230801 (2012).

\bibitem{Numata} K. Numata, A. Kemery, und J. Camp, Phys. Rev. Lett. \textbf{93,} 250602 (2004).  

\bibitem{Cole} G. D. Cole, W. Zhang, M. J. Martin, J. Ye,
M. Aspelmeyer, Nature Photonics \textbf{7,} 644--650 (2013).  

\bibitem{Yamamoto} K. Yamamoto, S. Miyoki, T. Uchiyama,
H. Ishitsuka, M. Ohashi, K. Kuroda, T. Tomaru, N. Sato, T. Suzuki,
T. Haruyama, A. Yamamoto, T. Shintomi, K. Numata, K. Waseda, K. Ito,
und K. Watanabe, Phys. Rev. D \textbf{74,} 022002 (2006). 

\bibitem{Granata} M. Granata, K. Craig, G. Cagnoli, C.
Carcy, W. Cunningham, J. Degallaix, R. Flaminio, D. Forest, M. Hart,
J.-S. Hennig, J. Hough, I. MacLaren, I. W. Martin, C. Michel, N. Morgado,
S. Otmani, L. Pinard, und S. Rowan, Opt. Lett. \textbf{38,} 5268--5271 (2013). 

\bibitem{Seel} S. Seel, R. Storz, G. Ruoso, J. Mlynek, und
S. Schiller, Physical Review Letters \textbf{78,} 4741--4744 (1997).

\bibitem{StorzDisseration} R. Storz, Dissertation, Univ. Konstanz (1998).

\bibitem{Storz} R. Storz, C. Braxmaier, K. Jack, O. Pradl,
und S. Schiller, Optics Letters \textbf{23,} 1031--1033 (1998). 

\bibitem{Antonini} P. Antonini, M. Okhapkin, E. Goklü, S.
Schiller, Physical Review A \textbf{71,} 050101, (2005). 

\bibitem{Mueller} C. Braxmaier, H. Müller, O. Pradl,
J. Mlynek, A. Peters, S. Schiller, Physical Review Letters \textbf{88,} 010401 (2001). 

\bibitem{Schiller} S. Schiller, C. Lammerzahl, H. Müller,
C. Braxmaier, S. Herrmann, A. Peters, Physical Review D \textbf{69,} 027504 (2004).  

\bibitem{Lyon1977} K. G. Lyon, G. L. Salinger, C. A. Swenson,
and G. K. White, J. Appl. Phys. \textbf{48,} 865 (1977).

\bibitem{McGuigan} D. F.~McGuigan, C. C. Lam, R. Q.
Gram, A. W. Hoffman, D. H. Douglass, H. W. Gutche, J. Low Temp. Phys. \textbf{30,} 621--629 (1978). 

\bibitem{Richard} J.-P.~Richard, J. J. Hamilton, Rev. Sci. Instr. \textbf{62,} 2375--2378 (1991). 

\bibitem{Kessler2012} T. Kessler, C. Hagemann, C. Grebing,
T. Legero, U. Sterr, and F. Riehle, Nature Photonics \textbf{6,} 687, (2012).

\bibitem{Sterr} U. Sterr, German patent DE 10 2011 015 489.2, (2011).

\bibitem{Drever} R. W. P. Drever, J. L. Hall, F. V. Kowalski, J. Hough, G. M. Ford, A. J. Munley, and H. Ward, Appl. Phys. B \textbf{31,} 97 (1983).

\bibitem{Smith1975} T.F. Smith, G.K. White, J. Phys. C \textbf{8,} 2031--2042 (1975).

\bibitem{Chen2012} Q.-F. Chen, A.Yu. Nevsky, and S. Schiller, Appl. Phys. B \textbf{107,} 679--683 (2012).

\end{thebibliography}
\end{document}